\documentclass[prl,aps,twocolumn,superscriptaddress,floatfix,showpacs,longbibliography]{revtex4-2}
\usepackage{graphicx}
\usepackage{dcolumn}
\usepackage{bm}
\usepackage{upgreek}
\usepackage{natbib}
\usepackage[normalem]{ulem}
\usepackage{amsmath,amssymb}
\usepackage{dsfont}
\usepackage[english]{babel}
\usepackage{color}

\begin{document}

\title{Single photon emitters with polarization and orbital angular momentum locking in monolayer semiconductors}

\author{Di Zhang}
\affiliation{Guangdong Provincial Key Laboratory of Quantum Engineering and
Quantum Materials, School of Physics and Telecommunication Engineering, South
China Normal University, Guangzhou 510006, China}
\affiliation{Guangdong-Hong Kong Joint Laboratory of Quantum Matter, Frontier
Research Institute for Physics, South China Normal University, Guangzhou
510006, China}
\author{Sha Deng}
\affiliation{Guangdong Provincial Key Laboratory of Quantum Engineering and
Quantum Materials, School of Physics and Telecommunication Engineering, South
China Normal University, Guangzhou 510006, China}
\affiliation{Guangdong-Hong Kong Joint Laboratory of Quantum Matter, Frontier
Research Institute for Physics, South China Normal University, Guangzhou
510006, China}
\author{Dawei Zhai}
\affiliation{Department of Physics, The University of Hong
Kong, Hong Kong, China}
\affiliation{HKU-UCAS Joint Institute of
Theoretical and Computational Physics at Hong Kong, China}
\author{Wang Yao}
\affiliation{Department of Physics, The University of Hong
Kong, Hong Kong, China}
\affiliation{HKU-UCAS Joint Institute of
Theoretical and Computational Physics at Hong Kong, China}
\author{Qizhong Zhu}
\email{qzzhu@m.scnu.edu.cn}
\affiliation{Guangdong Provincial Key Laboratory of Quantum Engineering and
Quantum Materials, School of Physics and Telecommunication Engineering, South
China Normal University, Guangzhou 510006, China}
\affiliation{Guangdong-Hong Kong Joint Laboratory of Quantum Matter, Frontier
Research Institute for Physics, South China Normal University, Guangzhou
510006, China}

\date{\today}

\begin{abstract}
  Excitons in monolayer transition metal dichalcogenide are endowed with intrinsic
  valley-orbit coupling between their center-of-mass motion and valley pseudospin. 
  When trapped in a confinement potential, e.g., generated by strain field, we find
  that intralayer excitons are valley and orbital
  angular momentum (OAM) entangled. By tuning trap profile and external magnetic
  field, one can engineer the exciton states at ground state, and realize a
  series of valley-OAM entangled states. We further show that the OAM of excitons can be
  transferred to emitted photons, and these novel
  exciton states can naturally serve as polarization-OAM locked single photon
  emitters, which under certain circumstance become polarization-OAM entangled, 
  highly tunable by strain trap and magnetic field. Our proposal
  demonstrates a novel scheme to generate polarization-OAM locked/entangled
  photons at nanoscale with high degree of integrability and tunability, pointing to exciting
  opportunities for quantum information applications.
\end{abstract}

{\maketitle}

Single photon emitters play a key role in quantum optics and quantum information technologies \cite{lounis_Singlephoton_2005,
buckley_Engineered_2012,aharonovich_Solidstate_2016,
senellart_Highperformance_2017,arakawa_Progress_2020}. 
Photons carrying orbital angular momentum (OAM) is of particular
importance for high-capacity quantum information processing. Tremendous effort has been made for generating photons with 
OAM in various platforms \cite{bomzon_Radially_2002,fickler_Quantum_2012,cai_Integrated_2012,lin_Nanostructured_2013,li_Holographic_2016,miao_Orbital_2016,
devlin_Arbitrary_2017,
shao_Spinorbit_2018,dorney_Controlling_2019,huang_Ultrafast_2020,zhang_Tunable_2020,
liu_Broadband_2021,chen_Bright_2021}, mostly relying on extra sophisticated phase transformers. 
More interestingly, the locking or entanglement between spin angular momentum (SAM) and OAM of photons, with further increased entanglement dimensionality, can also be
engineered using appropriate methods. For example, this entanglement has recently
been demonstrated by passing photons through geometric phase metasurface \cite{stav_Quantum_2018}, or 
using deterministically positioned quantum emitters combined with surface plasmon polaritons and
spiral gratings \cite{wu_Roomtemperature_2022}. 
On the whole, the schemes for generating polarization-OAM locked/entangled photons are rather limited and mostly rely on sophisticated 
experimental setup.

\begin{figure}
\centering
\includegraphics[width=0.95\linewidth]{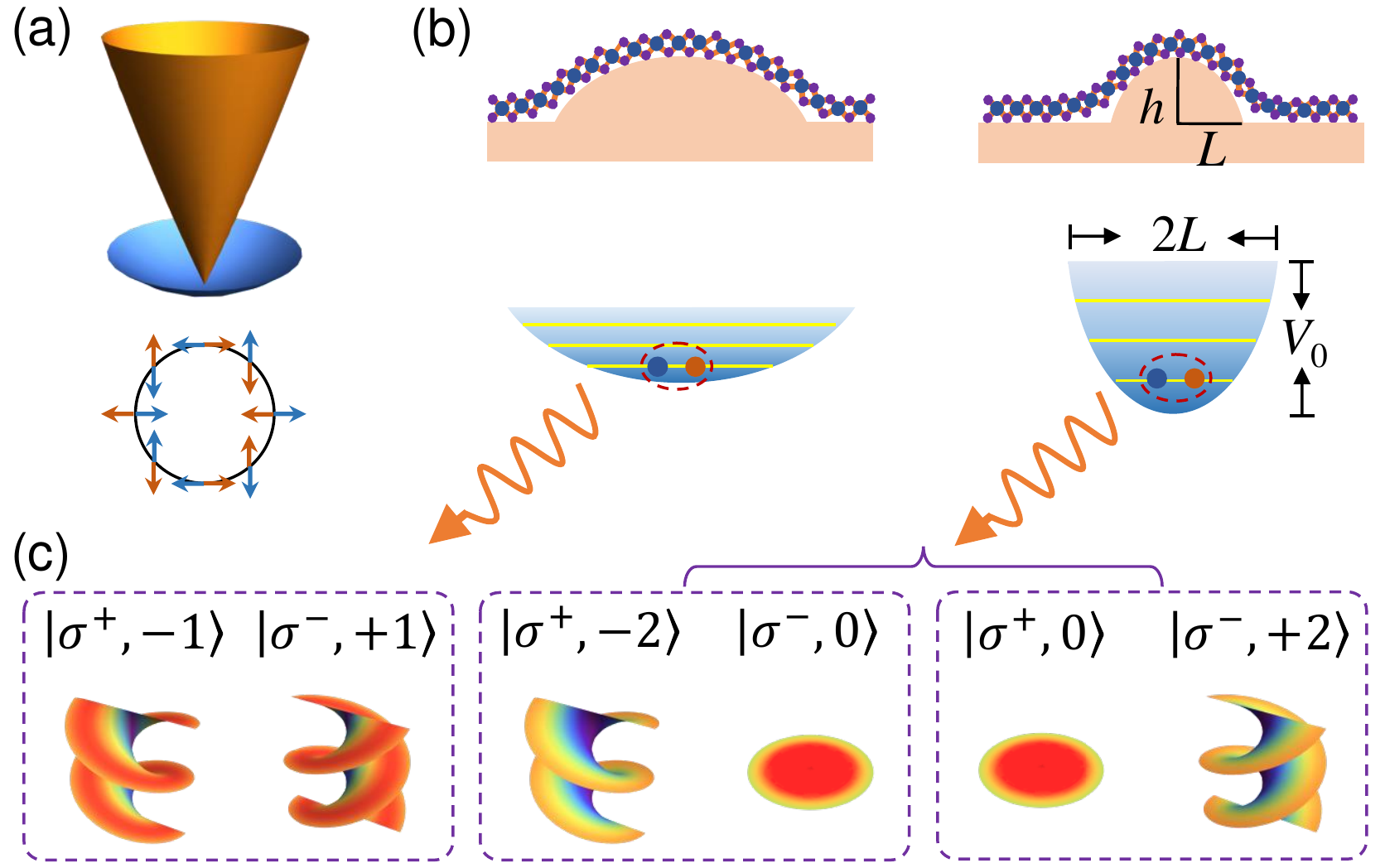}
\caption{(a) Upper: dispersions of intralayer exciton with valley-orbit coupling for $\delta=0$. Lower: textures of valley pseudospin denoted by arrows of
same color with corresponding branch. (b) Monolayer TMDs on
nanobubbles of different aspect ratios, generating shallow or tight strain traps for excitons. 
(c) Excitons at ground state of a shallow trap can recombine and emit photons in state $\{|\sigma^+,-1\rangle,|\sigma^-,1\rangle\}$;
 excitons at ground state of a tight trap can emit photons
either in state $\{|\sigma^+,-2\rangle,|\sigma^-,0\rangle\}$ or
$\{|\sigma^+,0\rangle,|\sigma^-,2\rangle\}$, tunable by external magnetic field. $\{|\sigma^+,l\rangle,|\sigma^-,l+2\rangle\}$
denotes photon state with OAM $l$ ($l+2$) in $\sigma^+$ ($\sigma^-$) polarization. Helical wavefronts illustrate the nonzero OAM
of these three different polarization-OAM locked/entangled photons. 
\label{fig1}}
\end{figure}

Nowadays, with the significant advances in transition metal dichalcogenide (TMD), 
people have realized that excitons in monolayer TMD as well as their heterostructures can serve
as an intriguing platform for single photon emitters \cite{turunen_Quantum_2022}. These systems enable selective control of exciton valley degree of
freedom through coupling with light polarization.
In fact, single photon emitters based on both monolayer and
heterobilayer TMDs are already realized \cite{he_Single_2015,koperski_Single_2015,chakraborty_Voltagecontrolled_2015,
tonndorf_Singlephoton_2015,
branny_Deterministic_2017,palacios-berraquero_Largescale_2017,linhart_Localized_2019,so_Electrically_2021}. 
With the versatile experimental techniques to engineer desired exciton trap using strain,
e.g., by placing TMD on nanopillar, nanobubble or nanosphere, the properties of single emitters based
on trapped excitons are highly controllable (see Fig. \ref{fig1}(b)). On the other hand,
in heterobilayers, moir\'e potentials can also provide an exciton trap,
which possibly realize a perfect array of quantum emitters based on moir\'e excitons \cite{yu_Moire_2017,baek_Highly_2020}.
TMD based single photon emitters have the advantage of high integrability and photon
extraction efficiency, which is especially desirable for quantum information applications. 

Till now, photons from these TMD based single photon emitters do not carry OAM, with only two SAM/polarization states. 
Interestingly, there is an unique property of intralayer exciton in monolayer TMD, i.e.,
coupling between exciton valley pseudospin with center-of-mass momentum \cite{yu_Dirac_2014,qiu_Nonanalyticity_2015}, originating
from electron-hole Coulomb exchange interaction. The valley-orbit coupling has chirality of two,
inducing entanglement between valley pseudospin and OAM of excitons. Since excitons in $+K$ ($-K$) valley
are interconvertable with photons of $\sigma^+$ ($\sigma^-$) polarization,
we find that, this valley-orbit coupling can be exploited to couple SAM and
OAM of emitted photons. The three degrees of freedom of photons, i.e., SAM, OAM and temporal profile are entangled consequently, 
reducing to
SAM-OAM entanglement if the temporal profile in two SAM channels are the same.
In the general case, we adopt the terminology ``polarization-OAM locking'' for rigorousness.
 In particular, the emitted photons by exciton ground state intrinsically carry nonzero OAM in either polarization component,
without the need of extra phase transformers.
 Specifically, we first show that when confined in a trap, e.g., 
generated by strain field, the exciton eigenstates are valley-OAM entangled, and also
 OAM of these excitons can be transferred to photons via radiative 
recombination.
So those excitons can naturally serve as single emitters of photons
with SAM and OAM locking/entanglement. 
Our proposal provides an appealing candidate for realizing
SAM-OAM locked/entangled single photon emitters,
and these emitters
inherit the advantage of generic TMD based emitters.
Our results also have important implications for
moir{\'e} trapped intralayer excitons in bilayer TMDs, and
we subsequently propose a realistic system as a polarization-OAM locked single photon emitter array 
based on trions in moir\'e potentials from twist hBN substrate.

The key ingredient of our proposal relies on the intrinsic valley-orbit coupling of intralayer
exciton in monolayer TMD \cite{yu_Dirac_2014,qiu_Nonanalyticity_2015}.
With electron-hole Coulomb exchange interaction,
the center-of-mass momentum of intralayer exciton is coupled with valley pseudospin.
The Hamiltonian describing the center-of-mass motion of intralayer excitons in momentum space
 reads
\begin{equation}
  \hat{H}_0 =  \frac{\hbar^2 Q^2}{2m}+ \beta Q+ \beta Q \left(\cos(2\phi_{\boldsymbol{Q}})\sigma_x 
   + \sin(2\phi_{\boldsymbol{Q}})\sigma_y\right)+\delta\sigma_z,
\end{equation}
where $\boldsymbol{Q}$ is the exciton center-of-mass momentum with magnitude $Q$, $\beta$ is the valley-orbit coupling strength, related with
the strength of Coulomb interaction, and $\delta$ is the valley splitting induced by external magnetic field,
known as the valley Zeeman effect \cite{aivazian_Magnetic_2015,stier_Exciton_2016}. $m$ is the exciton effective mass and Pauli matrices $\sigma_i$ ($i=x, y, z$) correspond to exciton valley pseudospin.
As shown in Fig. \ref{fig1}(a), this Hamiltonian gives rise to exciton dispersions with two branches,
$E_{\pm} (Q) = \hbar^2 Q^2/2m+\beta Q  \pm\sqrt{\beta^2 Q^2 +\delta^2}$,
with corresponding 
eigenstates $\chi_{\pm}\left(\boldsymbol{Q}\right)=\left(e^{-i\phi_{\boldsymbol{Q}}}, \pm e^{i\phi_{\boldsymbol{Q}}}\right)^T/\sqrt{2}$ when $\delta=0$.
 Without external magnetic field ($\delta=0$), the
upper branch is linear at small $Q$, while the lower branch is parabolic (see Fig. \ref{fig1}(a)).

The form of eigenstates implies different valley pseudospin textures locked with $\boldsymbol{Q}$ for two branches. In addition, when confined in an isotropic exciton trap, e.g., created by strain (see Fig. \ref{fig1}(b)),
the exciton eigenstates become valley-OAM entangled, as will be shown below.
Specifically, we consider a strain induced harmonic trap with finite depth, which should be a reasonable approximation
around the minimum of a generic isotropic confinement potential, and solve the exciton eigenstates
 taking into account valley-orbit coupling.
Such a confinement potential results from strain induced gap modulations, which has been shown to be the dominant effect in strained TMDs \cite{rostami_Theory_2015}.
 Generalization to other form of confinement potential is straightforward and the detailed form of trap
only quantitatively affects the results we present below, as long as it is isotropic. 
Then the center-of-mass motion of intralayer excitons is described by the Hamiltonian $\hat{H} = \hat{H}_0 + V_{\textrm{trap}} (R)$ with the
confinement potential
$V_{\textrm{trap}} (R) = m \omega^2 R^2/2 - V_0$.
Here both the trap frequency $\omega$ and trap depth $V_0$ can be easily tuned within
current experimental techniques
by designing the strain field profile. The confinement potential is assumed to be vanishing outside the trap radius
$L\equiv\sqrt{2V_0/m\omega^2}$. 
For $L>10$ nm, which is much larger than
monolayer exciton Bohr radius $a_B\sim 1$ nm, the internal degrees of freedom
of exciton can be neglected, with its center-of-mass motion describable by the above Hamiltonian.

The eigenvalues and eigenstates of this Hamiltonian can be first solved numerically in momentum space \cite{karr_Numerical_2010}, avoiding to express the valley-orbit coupling term in
coordinate space. Afterwards, exciton wave function in real space can be readily obtained
by Fourier transform.
For an isotropic strain trap, each pesudospin component of eigenstates can be characterized by definite
OAM with quantum number $l$.
The eigenstates in momentum space has the form
$\Psi(\boldsymbol{Q})=\left(\psi_K (Q) e^{i l \phi_{\boldsymbol{Q}}},
    \psi_{-K} (Q) e^{i (l + 2) \phi_{\boldsymbol{Q}}}\right)^T=\psi_K (Q)|K,l\rangle+\psi_{-K} (Q)|-K,l+2\rangle$,
hereafter denoted as $\left\{|K,l\rangle,|-K,l+2\rangle\right\}$, specifying the OAM $l$ ($l+2$) in $K$ ($-K$) valley
of exciton state.
Here the OAM of the two pseudospin components differ by $2$,
as required by total angular momentum conservation respected by valley-orbit coupling,
explicitly entangling valley pseudospin and OAM of excitons.
The exciton eigenvalues and eigenstates can be obtained by solving the following coupled equations (see Supplementary Material for further details),
\begin{widetext}
\begin{align}
E_l\left(\begin{array}{c}
    \psi_K (Q) \\
    \psi_{-K} (Q)
  \end{array}\right)=\left(\begin{array}{c}
    K_+(Q)\psi_K (Q)+\beta Q \psi_{-K} (Q)+\frac{1}{\left(2\pi\hbar\right)^2}\int d\boldsymbol{Q}'V\left(|\boldsymbol{Q}-\boldsymbol{Q}'|\right)
    \psi_K (Q') e^{-i l\left(\phi_{\boldsymbol{Q}}-\phi_{\boldsymbol{Q}'}\right)} \\
    K_-(Q)\psi_{-K} (Q)+\beta Q \psi_{K} (Q)+\frac{1}{\left(2\pi\hbar\right)^2}\int d\boldsymbol{Q}'V\left(|\boldsymbol{Q}-\boldsymbol{Q}'|\right)
    \psi_{-K} (Q') e^{-i(l+2)\left(\phi_{\boldsymbol{Q}}-\phi_{\boldsymbol{Q}'}\right)} 
  \end{array}\right),
\end{align}
\end{widetext}
where $K_{\pm}(Q)=\hbar^2 Q^2/2m+\beta Q\pm\delta$ and $V\left(|\boldsymbol{Q}-\boldsymbol{Q}'|\right)$ is the Fourier transform of confinement potential
$V(R)$.
The exciton wave function in coordinate space, obtained
by Fourier transform in polar coordinate, has the similar form $\Psi(\boldsymbol{R})=\left(\psi_K (R) e^{i l \phi_{\boldsymbol{R}}},\psi_{-K} (R) e^{i (l + 2) \phi_{\boldsymbol{R}}}\right)^T$, where
 $\boldsymbol{R}$ is center-of-mass coordinate of exciton.

\begin{figure}
\centering
\includegraphics[width=0.98\linewidth]{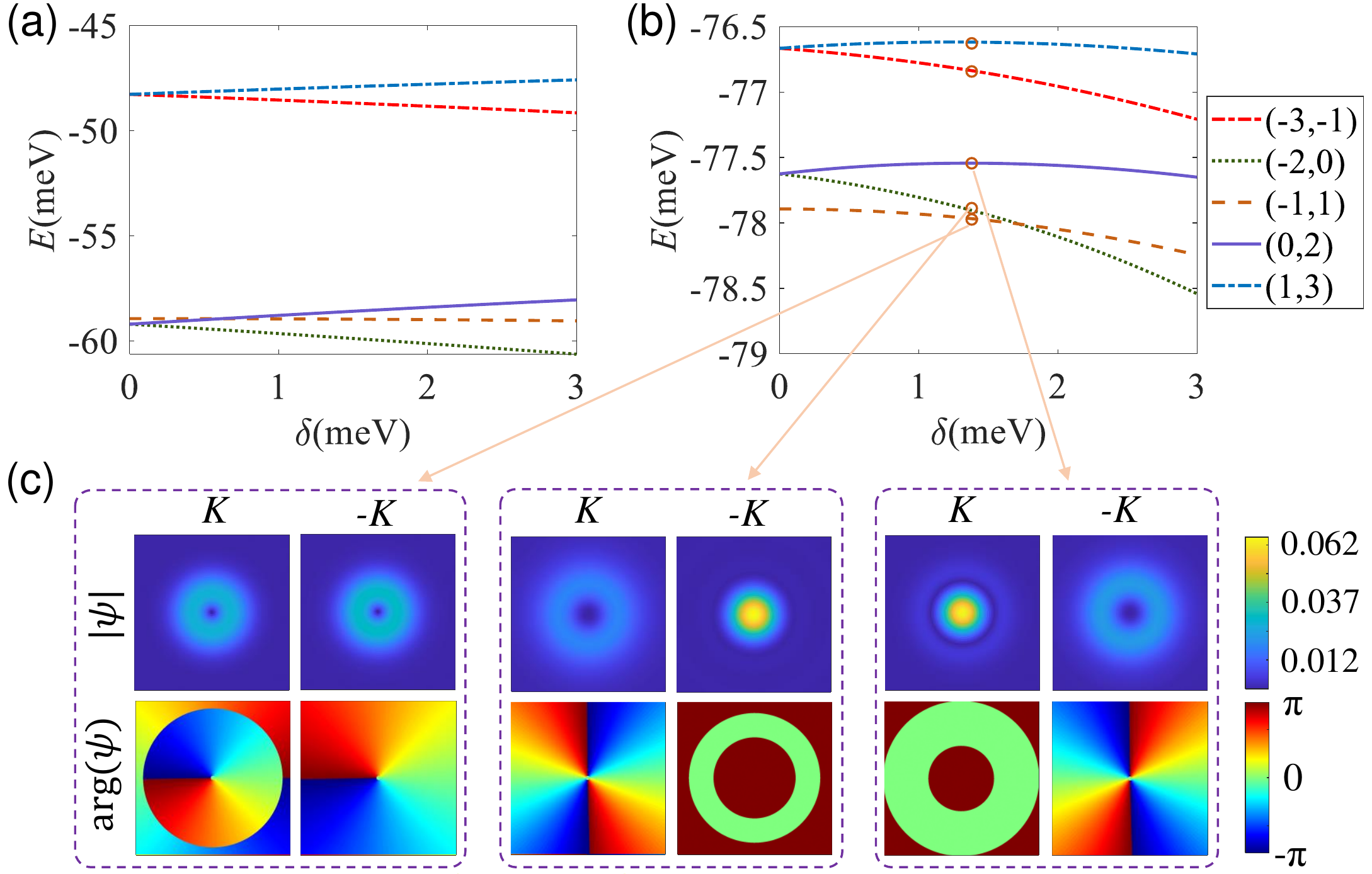}
\caption{Lowest $5$ exciton levels in a typical tight (a) and shallow trap (b). $(l,l+2)$ in the legend denotes the exciton state $\{|K,l\rangle,|-K,l+2\rangle\}$. (c) Exciton wave functions of the lowest $3$ energy levels
in (b) marked by small circles. Amplitude (phase) of wave function is shown in upper (lower) row, with
$K$ and $-K$ labelling the valley index. The square width is $70$ nm and the unit of $|\psi|$ is nm$^{-1}$.   \label{fig2}}
\end{figure}

By solving the above equations, we find that the ground state changes as the
dimensionless quantity $\tilde{\beta}=\beta/\left(\hbar\omega\sqrt{\hbar/m\omega}\right)$ exceeds a critical value
$\tilde{\beta}_c\approx 3\sim 4$, dependent on $V_0/\hbar\omega$.
When $\tilde{\beta}<\tilde{\beta}_c$, the ground state is two-fold degenerate, i.e., states
$\{|K,-2\rangle,|-K,0\rangle\}$ or $\{|K,0\rangle,|-K,2\rangle\}$. The degeneracy can be broken by
applying an out-of-plane magnetic field, i.e., $\delta\neq 0$. In contrast, when $\tilde{\beta}>\tilde{\beta}_c$,
the ground state is unique, i.e., state $\{|K,-1\rangle,|-K,1\rangle\}$. In the latter case, with the increase of magnetic field strength, 
the ground state 
changes from $\{|K,-1\rangle,|-K,1\rangle\}$ to $\{|K,-2\rangle,|-K,0\rangle\}$ or $\{|K,0\rangle,|-K,2\rangle\}$, 
depending on the direction of magnetic field. In a word, there are three possible non-degenerate ground states in the presence of
external magnetic field, each carrying nonzero OAM of excitons,
switchable by varying strain potential and magnetic field.

To be specific, we choose two trap profiles, each representative of one class of exciton levels, as shown in Fig. \ref{fig2}. The confinement potential profile created by strain is related with
the height $h$ and radius $L$ of the bubble (see Fig. \ref{fig1}(b)), which can be flexibly tuned via substrate
engineering. The maximum strain $\epsilon$ on the monolayer TMD depends on the aspect ratio of bubble through $\epsilon\sim h^2/L^2$ \cite{chirolli_Straininduced_2019}, 
and consequently the strain induced trap depth $V_0$ is determined as $V_0\sim\gamma h^2/L^2$,
with $\gamma\approx 30\sim 40$ meV per $1\%$ of strain for TMD. For realistic strain $\epsilon\lesssim 5\%$, a trap depth $V_0\sim O(100)$ meV
is readily achievable. Within the harmonic trap model, the trap frequency is therefore determined by $\omega=\sqrt{2V_0/mL^2}$.
Figure \ref{fig2}(a) corresponds to a tight trap, with $V_0=80$ meV, $L=10$ nm ($\hbar\omega\approx10.5$ meV), and the dimensionless $\tilde{\beta}=3.3$, below the critical value. 
Figure \ref{fig2}(b) corresponds to a shallow trap, 
with $V_0=80$ meV, $L=100$ nm ($\hbar\omega\approx1.1$ meV), and $\tilde{\beta}=10.5$, above the critical value. The bare parameters are  
$\beta=0.9$ eV$\cdot$\AA, and exciton effective mass $m=1.1 m_e$,
with $m_e$ being electron mass \cite{qiu_Nonanalyticity_2015}.
The lowest $5$ exciton levels in each scenario along with their change with valley Zeeman energy $\delta$ are shown in Fig. \ref{fig2}.
The magnitude and phase of wave functions for the lowest three exciton levels are also shown in Fig. \ref{fig2}(c), demonstrating the
above mentioned valley-OAM entanglement.

The valley and OAM
entanglement of excitons implies novel optical selection rules when they are coupled with light.
Microscopically, we start from the light-matter interaction Hamiltonian,
$\hat{H}_I=-e\boldsymbol{\mathcal{A}}\cdot\hat{\boldsymbol{p}}/m$, and calculate the
exciton-light coupling matrix element $\mathcal{T}=\langle\Upsilon|\hat{H}_I|0\rangle$, where 
$\Upsilon(\boldsymbol{R},\boldsymbol{r})=\Psi(\boldsymbol{R})
\otimes\Phi(\boldsymbol{r})$ is the exciton wave function composed of both center-of-mass part $\Psi(\boldsymbol{R})$ as 
calculated above and internal part $\Phi(\boldsymbol{r})$. 
Recently, it was shown that the OAM of photon in twisted light can be transferred
to the center-of-mass OAM of exciton \cite{grass_Twodimensional_2022}.
We adopt similar formalism and find that those excitons with valley and OAM entanglement
 will couple with
photons with polarization-OAM locking. 

$\Psi(\boldsymbol{R})$ have
two components $\Psi_{\pm}(\boldsymbol{R})$ corresponding to $\pm K$ valley, and $\Phi(\boldsymbol{r})$
is assumed to be in the $s$-wave state of electron-hole relative motion.  
A general vector potential of light field can be decomposed as $\boldsymbol{\mathcal{A}}=\boldsymbol{\hat{\varepsilon}}_+A_+(\boldsymbol{R})+
\boldsymbol{\hat{\varepsilon}}_-A_-(\boldsymbol{R})$,
with $\boldsymbol{\hat{\varepsilon}}_\pm$ being the unit vector of $\sigma^\pm$ polarization.
The matrix element can be separated accordingly, $\mathcal{T}=\mathcal{T}_{+}+\mathcal{T}_{-}$, with the $+$ ($-$) component 
contributed by coupling between $K$ ($-K$) valley of excitons and $\boldsymbol{\hat{\varepsilon}}_+$
($\boldsymbol{\hat{\varepsilon}}_-$) polarization of vector potential.
The matrix element is given by (see Supplementary Material for detailed derivation)
\begin{align}
\mathcal{T}_{\pm} & \propto\int d\boldsymbol{R}A_\pm(\boldsymbol{R})\Psi_{\pm}^{*}(\boldsymbol{R})
\int d\boldsymbol{r}\Phi^{*}(\boldsymbol{r})\int d\boldsymbol{k}e^{i\boldsymbol{k}\cdot\boldsymbol{r}}{p}_{\pm}^{\mathrm{vc}}\nonumber\\
& =\mathcal{N}\int d\boldsymbol{R}A_\pm(\boldsymbol{R})\Psi_{\pm}^{*}(\boldsymbol{R}),
\end{align}
where ${p}_{\pm}^{\mathrm{vc}}$
is the dipole moment between conduction and valence band in $\pm K$ valley, with dependence on electron momentum neglected, and $\mathcal{N}$ is a coefficient quantifying the transition amplitude of $s$-wave exciton.
So the transition matrix element is proportional to the overlap
between the exciton center-of-mass wave function and laser vector potential, related with the excitation rate $\eta$ of a laser beam.

Clearly, for rotationally symmetric excitation field $\boldsymbol{\mathcal{A}}$ with definite OAM, 
the matrix element does not vanish only when the exciton OAM coincides with
the OAM of excitation light. So exciton state $\{|K,l\rangle,|-K,l+2\rangle\}$ will couple with photon state
$\{|\sigma^+,l\rangle,|\sigma^-,l+2\rangle\}$, characterized by vector potential
$\mathcal{A}_+ (\boldsymbol{R}) = \boldsymbol{\hat{\varepsilon}}_+ {A}_{l}(r) e^{i l \phi_{\boldsymbol{R}}}$,
$\mathcal{A}_- (\boldsymbol{R}) = \boldsymbol{\hat{\varepsilon}}_- {A}_{l+2}(r) e^{i (l+2) \phi_{\boldsymbol{R}}}$ (the overall out-of-plane
part $e^{i q_z z}$ is omitted here).
This indicates the OAM transfer from exciton to photon and vice versa, 
and the entanglement between exciton valley and OAM
will give rise to locking between polarization and OAM of single photons. Note that
temporal profiles of excitons in two valleys are generally different, due to different radiative lifetimes, 
and therefore the emitted photons have entanglement between SAM, OAM and temporal wave function.
In the case of equal valley radiative lifetime, there exists quantum entanglement between SAM and OAM of photons; otherwise, 
the more accurate terminology ``polarization-OAM locking'' is used in general cases.
 To quantitatively determine to what extent ``locking'' becomes ``entanglement'',
we have calculated the radiative lifetimes of lowest 3 exciton states in Fig. \ref{fig2}(c) for each valley \cite{wang_Radiative_2016},
which are $(\tau_K, \tau_{-K})\approx(225.5\tau_0, 138.8\tau_0)$, $(902.6\tau_0, 1.4\tau_0)$ and $(6.6\tau_0, 1013.7\tau_0)$, respectively.
With $\tau_0\approx 0.2$ ps being the radiative lifetime of unconfined excitons at vanishing center-of-mass momentum, the radiative lifetimes of these
trapped excitons are on the order of $10\sim 100$ ps. The lifetime can be further prolonged (shortened) by choosing a tighter (shallower) trap.
 A series of exciton levels can be exploited to serve
as single photon sources with desired OAM. For example, the three exciton states in Fig. \ref{fig2}(c)
will emit three types of polarization-OAM locked/entangled photons illustrated in Fig. \ref{fig1}(c).
Interestingly, photon state
$\{|\sigma^+,-1\rangle,|\sigma^-,1\rangle\}$ will become polarization-OAM entangled
in the absence of magnetic field.

Recently, there has been growing interest in manipulating excitons in TMD materials
with twisted light \cite{simbulan_Selective_2021,peng_Tailoring_2022,kesarwani_Control_2022}, 
i.e., light carrying nonzero OAM. Here using twisted light in Laguerre-Gaussian (LG) modes with matched OAM, one can directly excite
those exciton states with nonzero OAM. For example, exciton state $\{|K,-1\rangle,|-K,1\rangle\}$ at ground state in Fig. \ref{fig2}(c)
will be optically dark when probed with fundamental Gaussian mode ($l=0$), but be bright
for light of LG$_{l=\mp1,p=0}$ mode in $\sigma^{\pm}$ polarization.

\begin{figure}
\centering
\includegraphics[width=0.98\linewidth]{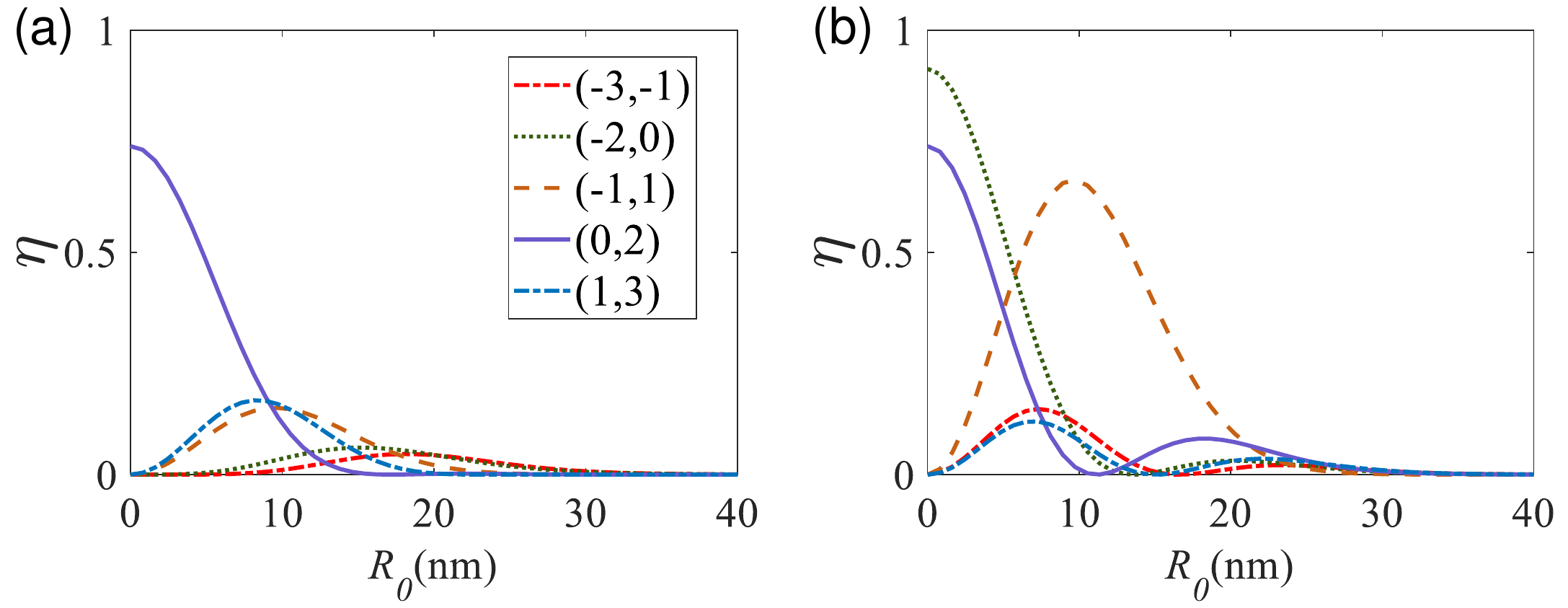}
\caption{Excitation rate (in arbitrary unit) of a laser field with Gaussian profile for the lowest $5$ exciton states marked by small circles in Fig. \ref{fig2}(b). The excitation laser is
 $\sigma^+$ polarized in (a) and linearly polarized in (b). $R_0$ is the distance between excitation center and trap center.
 \label{fig3}}
\end{figure}

On the other hand, if one excites those excitons carrying OAM using laser in fundamental Gaussian mode,
the excitation rate can still be nonzero if the excitation center does not coincide with the trap center.
In Fig. \ref{fig3}, we plot the excitation rate $\eta\equiv|\int d\boldsymbol{R}\sum_\pm A_\pm(\boldsymbol{R})\Psi_{\pm}^{*}(\boldsymbol{R})|^2$ 
as a function of the distance $R_0$ between excitation center and trap center for the 5 exciton states in Fig. \ref{fig2}(b).
We choose a Gaussian excitation field $\mathcal{G}=\exp(-(\boldsymbol{R}-\boldsymbol{R}_0)^2/w^2)$, with
$A_+=\mathcal{G}$, $A_-=0$ in Fig. \ref{fig3}(a) and $A_{\pm}=\mathcal{G}$ in Fig. \ref{fig3}(b).
The Gaussian wave width $w$ is set to be $10$ nm, achievable with tip-enhanced techniques \cite{park_Hybrid_2016,qiu_3D_2018,gao_Selective_2021}. Clearly,
there is non-vanishing excitation rate for off-center laser excitation,
since in this case the laser beam does not carry definite OAM with respect to trap center.

The main results above, including valley-OAM entangled excitons and the resulted polarization-OAM locked photons, are applicable to any isotropic exciton trap, provided that the trap size is much larger
than exciton Bohr radius.
To experimentally observe the above predicted phenomena, one has to separate the strain trapped exciton states
from defect trapped exciton states, as defects also play an important role in current demonstration of single photon emitters
in monolayer TMD. 
In principle, these two types of exciton levels can be carefully resolved by photon energy in Photonluminance (PL), since their energy
windows are generally different.

These exciton
states with nonzero OAM and valley-OAM entanglement also exist in moir\'e traps provided by twisted TMD heterobilayers. 
Recently, it is discovered that
strong lattice reconstruction usually occurs in bilayer TMDs with large moir\'e size \cite{weston_Atomic_2020,li_Imaging_2021,naik_Intralayer_2022}, 
invaliding the applicability of continuum model \cite{wu_Topological_2017},
previously used to describe moir\'e intralayer excitons. This poses a challenge to the realization of single photon emitter array based on moir\'e excitons,
since negligible hybridization of exciton wave packets between neighbouring trap minima is required, which is however hindered by
the limited moir\'e size available (see Supplementary Material for details).
This obstacle motivates us to propose another system, i.e., intralayer trions in moir\'e potentials generated by twisted multilayer hBN substrate \cite{zhao_Universal_2021},
where lattice reconstruction should be weak even for large moir\'e size. 

\begin{figure}
\centering
\includegraphics[width=0.98\linewidth]{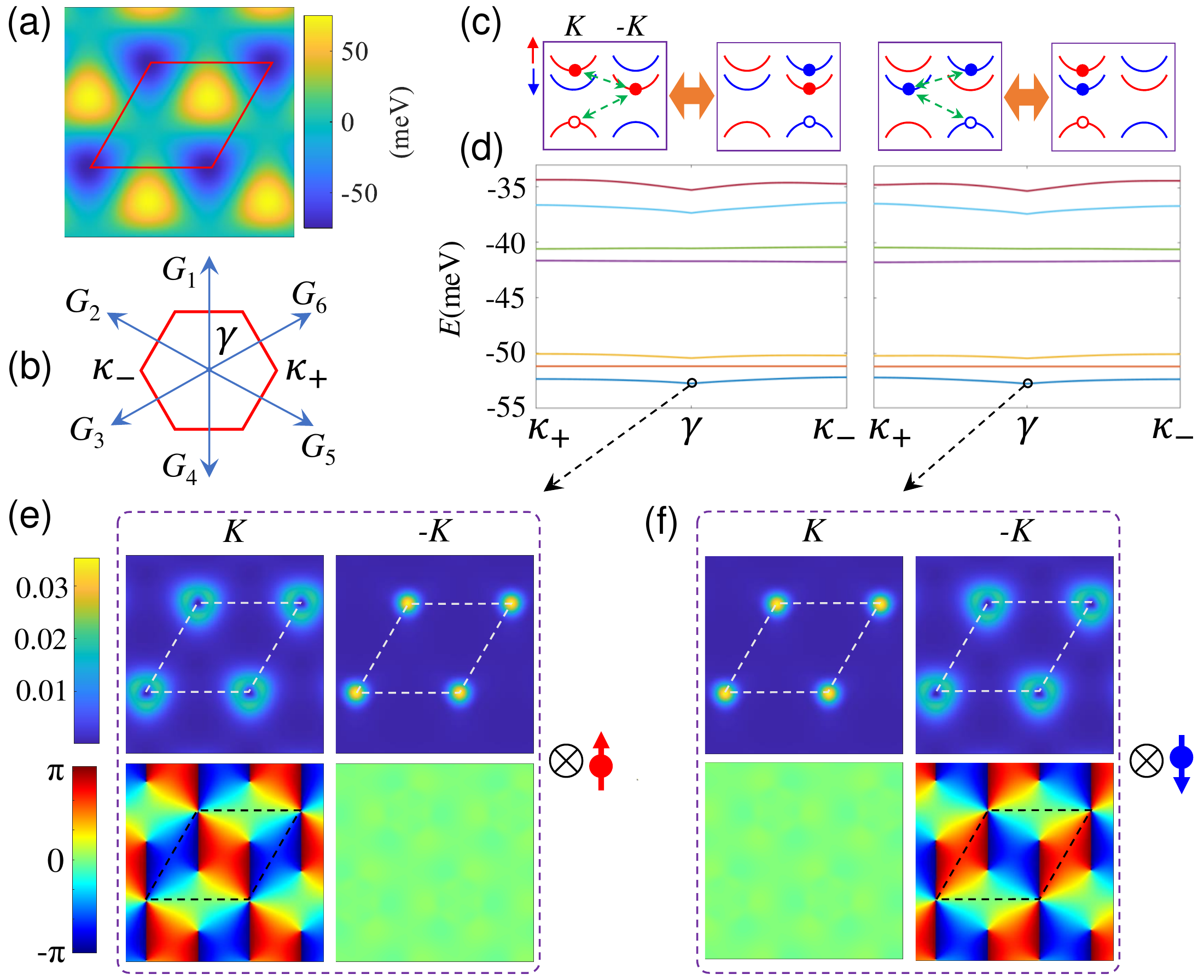}
\caption{(a) Moir\'e potential for intralayer trions generated by twisted hBN substrate. The moir\'e period is $a_m\approx 28.7$ nm corresponding
to $\theta=0.5^{\circ}$.
(b) Moir\'e Brillouin zone and six reciprocal lattice vectors used to construct moir\'e potential in (a). Trion momentum is measured relative to 
$\pm K$ valley of monolayer TMD.
(c) Spin-valley configurations of four types of trions, where the former and latter two configurations are related by valley-orbit coupling
 (indicated by orange double arrows).
(d) Moir\'e minibands for two groups of trions illustrated in (c). Left (right) panel corresponds to the case of spin up (down) excess electron.
(e),(f) Trion wave functions at the ground state marked by black circles in minibands of (d), with additional arrows after ``$\otimes$'' indicating spin of excess electron.
The amplitude (phase) of trion wave function is shown in upper (lower) row.
 \label{fig4}}
\end{figure}

The registry dependent electrical polarization at the twisted interface in the hBN substrate can generate an electrostatic superlattice potential for charged carriers \cite{zhao_Universal_2021},
and a negative trion in monolayer TMD placed on twisted hBN substrate experiences a moir\'e potential, modelled as
$V_{\mathrm{moir\acute{e}}}=\sum_{j=1}^6 V_j \exp(i\boldsymbol{G}_j\cdot\boldsymbol{R})$,
where $\boldsymbol{G}_j$ is the moir\'e reciprocal lattice vector with length $G$ (see Fig. \ref{fig4}(b)), and $V_j=V\exp(-Gnd)\exp(i\phi\times(-1)^j)$ characterizes
the strength of moir\'e potential, tunable by the twist angle $\theta$ between multilayer hBN interface and layer number $n$.
First-principles calculations \cite{zhao_Universal_2021} suggest $V=-19.5$ meV, $\phi=-\pi/2$, and $\boldsymbol{G}_1=(0,4\pi/\sqrt{3}a_m)$, with moir\'e period $a_m=a_0/\theta$. $d=0.3$ nm and $a_0=0.25$ nm are the hBN monolayer thickness and lattice constant, respectively.
We choose $\theta=0.5^{\circ}$ and layer number $n=4$, which should be large enough
for suppression of lattice reconstruction.
The trion Hamiltonian with this type of moir\'e potential reads \cite{yu_Dirac_2014},
\begin{align}
  \hat{H}_{\mathrm{trion}} = & \frac{\hbar^2 Q^2}{2m_t}+\beta Q + \beta Q \cos(2\phi_{\boldsymbol{Q}})\sigma_x+\beta Q \sin(2\phi_{\boldsymbol{Q}})\sigma_y \nonumber\\
  & +\frac{\Delta}{2}\left(\sigma_z s_z+1\right)+V_{\mathrm{moir\acute{e}}},
\end{align}
where $m_t\approx 1.6 m_e$ is the trion effective mass, $s_z$ is the spin of excess electron, and
$\Delta\approx 6$ meV characterizes the Coulomb exchange interaction between excess electron and electron-hole pair \cite{yu_Dirac_2014} (see green arrows in Fig. \ref{fig4}(c)), playing a similar role as valley Zeeman splitting.
 The trion moir\'e bands are shown
in Fig. \ref{fig4}(d), for two different spin polarizations of excess electron. The Bloch wave function
at ground state is also shown, which clearly exhibits the phase winding around moir\'e potential minima,
corresponding to state $\{|K,-2\rangle,|-K,0\rangle\}$ ($\{|K,0\rangle,|-K,2\rangle\}$) in Fig. \ref{fig4}(e) (Fig. \ref{fig4}(f)),
and should couple with photon state $\{|\sigma^+,-2\rangle,|\sigma^-,0\rangle\}$ ($\{|\sigma^+,0\rangle,|\sigma^-,2\rangle\}$).
Note that the character of these photons from single emitter array we proposed here is also conditioned on the spin polarization of excess electron.

In summary, we have proposed that, due to the intrinsic valley-orbit coupling, 
trapped intralayer excitons in monolayer TMD by a potential, e.g., provided by strain field, can serve as
single photon emitters with polarization-OAM locking/entanglement. The character of emitted photons by excitons at ground state
can be tuned by varying trap frequency and external magnetic field, and generally carry nonzero OAM.
The valley-OAM entangled exciton states can be directly probed by twisted light with matched OAM,
or off-center laser excitation with Gaussian profile. We also propose that, benefiting from moir\'e potential created by twisted
multilayer hBN substrate, with large moir\'e size and weak lattice reconstruction,
intralayer trions trapped by this moir\'e potential can form an array of single photon emitters with polarization-OAM locking.
Our work demonstrates a novel scheme to realize polarization-OAM locked/entangled single photon emitters and their array, 
with the advantage of high controllability and integrability, pointing to promising quantum information applications. 

We are grateful to Xu-Chen Yang, Chengxin Xiao and Hongyi Yu for valuable discussions.
D.Z., S.D. and Q.Z. are supported by the National Key Research and Development Program of China (Grant No. 2022YFA1405304),
National Natural
Science Foundation of China (Grant No. 12004118), and the Guangdong Basic and
Applied Basic Research Foundation (Grants No. 2020A1515110228 and No.
2021A1515010212). 
D.Z. and W.Y. acknowledge support by Research Grant Council of Hong Kong SAR (AoE/P-701/20, HKU SRFS2122-7S05) and Croucher Foundation.

\end{document}


\title{Supplemental Material for ``Single photon emitters with polarization and
orbital angular momentum locking in monolayer semiconductors''}

\author{Di Zhang}
\affiliation{Guangdong Provincial Key Laboratory of Quantum Engineering and
Quantum Materials, School of Physics and Telecommunication Engineering, South
China Normal University, Guangzhou 510006, China}
\affiliation{Guangdong-Hong Kong Joint Laboratory of Quantum Matter, Frontier
Research Institute for Physics, South China Normal University, Guangzhou
510006, China}
\author{Sha Deng}
\affiliation{Guangdong Provincial Key Laboratory of Quantum Engineering and
Quantum Materials, School of Physics and Telecommunication Engineering, South
China Normal University, Guangzhou 510006, China}
\affiliation{Guangdong-Hong Kong Joint Laboratory of Quantum Matter, Frontier
Research Institute for Physics, South China Normal University, Guangzhou
510006, China}
\author{Dawei Zhai}
\affiliation{Department of Physics, The University of Hong
Kong, Hong Kong, China}
\affiliation{HKU-UCAS Joint Institute of
Theoretical and Computational Physics at Hong Kong, China}
\author{Wang Yao}
\affiliation{Department of Physics, The University of Hong
Kong, Hong Kong, China}
\affiliation{HKU-UCAS Joint Institute of
Theoretical and Computational Physics at Hong Kong, China}
\author{Qizhong Zhu}
\email{qzzhu@m.scnu.edu.cn}
\affiliation{Guangdong Provincial Key Laboratory of Quantum Engineering and
Quantum Materials, School of Physics and Telecommunication Engineering, South
China Normal University, Guangzhou 510006, China}
\affiliation{Guangdong-Hong Kong Joint Laboratory of Quantum Matter, Frontier
Research Institute for Physics, South China Normal University, Guangzhou
510006, China}

\date{\today}
\maketitle

\section{I. Numerical procedure of solving exciton energy levels confined in
a trap}

The non-analytic form of valley-orbit coupling makes it inconvenient
to solve the exciton energy levels in coordinate space. In contrast,
it is much more convenient to solve in momentum space, where the eigen-equations
become
\begin{equation}
E_{l}\left(\begin{array}{c}
\Psi_{K}(\boldsymbol{Q})\\
\Psi_{-K}(\boldsymbol{Q})
\end{array}\right)=\left(\begin{array}{c}
K_{+}({Q})\Psi_{K}(\boldsymbol{Q})+\beta Qe^{-i2\phi_{\boldsymbol{Q}}}\Psi_{-K}(\boldsymbol{Q})+\frac{1}{\left(2\pi\hbar\right)^{2}}\int d\boldsymbol{Q}'V\left(|\boldsymbol{Q}-\boldsymbol{Q}'|\right)\Psi_{K}(\boldsymbol{Q}')\\
K_{-}({Q})\Psi_{-K}(\boldsymbol{Q})+\beta Qe^{i2\phi_{\boldsymbol{Q}}}\Psi_{K}(\boldsymbol{Q})+\frac{1}{\left(2\pi\hbar\right)^{2}}\int d\boldsymbol{Q}'V\left(|\boldsymbol{Q}-\boldsymbol{Q}'|\right)\Psi_{-K}(\boldsymbol{Q}')
\end{array}\right),
\end{equation}
with $K_{\pm}(Q)=\hbar^{2}Q^{2}/2m+\beta Q\pm\delta$ and $V\left(|\boldsymbol{Q}-\boldsymbol{Q}'|\right)$
being the Fourier transform of confinement potential $V(R)$. For
the isotropic confinement potential $V_{\mathrm{trap}}(R)=m\omega^{2}R^{2}/2-V_{0}$
adopted in the main text, the exciton wave functions can be classified
by orbital angular momentum $l$, and the eigenstates have the form
\begin{equation}
\Psi(\boldsymbol{Q})=\left(\begin{array}{c}
\psi_{K}(Q)e^{il\phi_{\boldsymbol{Q}}}\\
\psi_{-K}(Q)e^{i(l+2)\phi_{\boldsymbol{Q}}}
\end{array}\right),
\end{equation}
with the radial part of wave function $\psi_{K/-K}(Q)$ only dependent
on $Q=|\boldsymbol{Q}|$. Plugging the above wave function into the eigenvalue
equation, one arrives at
\begin{eqnarray}
E_{l}\left(\begin{array}{c}
\psi_{K}(Q)\\
\psi_{-K}(Q)
\end{array}\right) & = & \left(\begin{array}{c}
K_{+}(Q)\psi_{K}(Q)+\beta Q\psi_{-K}(Q)+\frac{1}{\left(2\pi\hbar\right)^{2}}\int d\boldsymbol{Q}'V\left(|\boldsymbol{Q}-\boldsymbol{Q}'|\right)\psi_{K}(Q')e^{-il\left(\phi_{\boldsymbol{Q}}-\phi_{\boldsymbol{Q}'}\right)}\\
K_{-}(Q)\psi_{-K}(Q)+\beta Q\psi_{K}(Q)+\frac{1}{\left(2\pi\hbar\right)^{2}}\int d\boldsymbol{Q}'V\left(|\boldsymbol{Q}-\boldsymbol{Q}'|\right)\psi_{-K}(Q')e^{-i(l+2)\left(\phi_{\boldsymbol{Q}}-\phi_{\boldsymbol{Q}'}\right)}
\end{array}\right)\nonumber \\
 & = & \left(\begin{array}{c}
K_{+}(Q)\psi_{K}(Q)+\beta Q\psi_{-K}(Q)+\frac{1}{\left(2\pi\hbar\right)^{2}}\int Q'dQ'd\phi_{\boldsymbol{Q}'}V\left(|\boldsymbol{Q}-\boldsymbol{Q}'|\right)\psi_{K}(Q')e^{-il\left(\phi_{\boldsymbol{Q}}-\phi_{\boldsymbol{Q}'}\right)}\\
K_{-}(Q)\psi_{-K}(Q)+\beta Q\psi_{K}(Q)+\frac{1}{\left(2\pi\hbar\right)^{2}}\int Q'dQ'd\phi_{\boldsymbol{Q}'}V\left(|\boldsymbol{Q}-\boldsymbol{Q}'|\right)\psi_{-K}(Q')e^{-i(l+2)\left(\phi_{\boldsymbol{Q}}-\phi_{\boldsymbol{Q}'}\right)}
\end{array}\right)\nonumber\\
 & = & \left(\begin{array}{c}
K_{+}(Q)\psi_{K}(Q)+\beta Q\psi_{-K}(Q)+\frac{1}{\left(2\pi\hbar\right)^{2}}\int Q'dQ'V_{l}\left(Q,Q'\right)\psi_{K}(Q')\\
K_{-}(Q)\psi_{-K}(Q)+\beta Q\psi_{K}(Q)+\frac{1}{\left(2\pi\hbar\right)^{2}}\int Q'dQ'V_{l+2}\left(Q,Q'\right)\psi_{-K}(Q')
\end{array}\right), 
\label{eq:Eigen}
\end{eqnarray}
where
\begin{eqnarray}
V_{l}\left(Q,Q'\right) & = & \int d\phi_{\boldsymbol{Q}'}V\left(|\boldsymbol{Q}-\boldsymbol{Q}'|\right)e^{-il\left(\phi_{\boldsymbol{Q}}-\phi_{\boldsymbol{Q}'}\right)}\nonumber\\
 & = & \int d\phi_{\boldsymbol{Q}'}V\left(Q,Q',\phi_{\boldsymbol{Q}}-\phi_{\boldsymbol{Q}'}\right)e^{-il\left(\phi_{\boldsymbol{Q}}-\phi_{\boldsymbol{Q}'}\right)}.
\end{eqnarray}
In the last line of Eq. \ref{eq:Eigen}, the eigenvalue equation has
been reduced to a coupled one-dimensional integral equations, which
can be numerically solved by discretization in $Q$ space. By choosing
different $l$ and solving the above equations, one obtains a series
of exciton energy levels for fixed $l$. By comparing the energy levels
of different $l$, one can determine the ground state, first excited
state, {\it et al}.

\section{II. Exciton-light coupling matrix element}

With the light-matter interaction Hamiltonian $H_{I}=-e\boldsymbol{\mathcal{A}}\cdot\hat{\boldsymbol{p}}/m$,
one can calculate the exciton-light coupling matrix element $\mathcal{T}=\langle\Upsilon|H_{I}|0\rangle$,
where $\Upsilon$ is the exciton wave function. Taking into account
both center-of-mass and relative motion of excitons, the exciton wave
function can be factorized to $\Upsilon_{\pm}(\boldsymbol{R},\boldsymbol{r})=\Psi_{\pm}(\boldsymbol{R})\otimes\Phi(\boldsymbol{r})$.
The transition matrix element for $\pm K$ valley can be calculated
as \cite{grass_Twodimensional_2022}
\begin{eqnarray}
\mathcal{T_{\pm}} & = & \langle\Upsilon_{\pm}|H_{I}|0\rangle\nonumber\\
 & = & \iint d\boldsymbol{R}d\boldsymbol{r}\iint d\boldsymbol{Q}d\boldsymbol{k}\langle\Upsilon_{\pm}|\boldsymbol{R},\boldsymbol{r}\rangle\langle\boldsymbol{R},\boldsymbol{r}|\boldsymbol{Q},\boldsymbol{k}\rangle\langle\boldsymbol{Q},\boldsymbol{k}|H_{I}|0\rangle\nonumber\\
 & = & \frac{1}{S^{2}}\iint d\boldsymbol{R}d\boldsymbol{r}\iint d\boldsymbol{Q}d\boldsymbol{k}\Psi_{\pm}^{*}(\boldsymbol{R})\Phi^{*}(\boldsymbol{r})e^{i\left(\boldsymbol{Q}\cdot\boldsymbol{R}+\boldsymbol{k}\cdot\boldsymbol{r}\right)}\langle\boldsymbol{k}-\frac{\boldsymbol{Q}}{2}|H_{I}|\boldsymbol{k}+\frac{\boldsymbol{Q}}{2}\rangle_{\pm},
\end{eqnarray}
where the completeness relation $\iint d\boldsymbol{R}d\boldsymbol{r}|\boldsymbol{R},\boldsymbol{r}\rangle\langle\boldsymbol{R},\boldsymbol{r}|=\iint d\boldsymbol{Q}d\boldsymbol{k}|\boldsymbol{Q},\boldsymbol{k}\rangle\langle\boldsymbol{Q},\boldsymbol{k}|=1$
has been inserted. $\langle\boldsymbol{Q},\boldsymbol{k}|H_{I}|0\rangle_{\pm}=\langle\boldsymbol{k}-\frac{\boldsymbol{Q}}{2}|H_{I}|\boldsymbol{k}+\frac{\boldsymbol{Q}}{2}\rangle_{\pm}$
is related with electron interband transition from Bloch state $\psi_{v,\boldsymbol{k}-\frac{\boldsymbol{Q}}{2}}(\boldsymbol{r})$
in valence band to state $\psi_{c,\boldsymbol{k}+\frac{\boldsymbol{Q}}{2}}(\boldsymbol{r})$
in conduction band in $\pm K$ valley. Since $\psi_{c/v,\boldsymbol{k}}(\boldsymbol{r})=e^{i\boldsymbol{k}\cdot\boldsymbol{r}}u_{c/v,\boldsymbol{k}}(\boldsymbol{r})$,
with $u_{c/v,\boldsymbol{k}}(\boldsymbol{r})$ being the periodic
part of Bloch wave function in conduction/valence band,
\begin{eqnarray}
\langle\boldsymbol{k}-\frac{\boldsymbol{Q}}{2}|H_{I}|\boldsymbol{k}+\frac{\boldsymbol{Q}}{2}\rangle_{\pm} & = & \frac{ie\hbar}{m}\int d\boldsymbol{r}e^{i\boldsymbol{Q}\cdot\boldsymbol{r}}u_{v,\boldsymbol{k}-\frac{\boldsymbol{Q}}{2}}^{*}\left(\boldsymbol{r}\right)A_{\pm}\left(\boldsymbol{r}\right)\boldsymbol{\hat{\varepsilon}}_{\pm}\cdot\boldsymbol{\nabla}u_{c,\boldsymbol{k}+\frac{\boldsymbol{Q}}{2}}\left(\boldsymbol{r}\right)\nonumber\\
 & \approx & \frac{ie\hbar}{m}\sum_{i}A_{\pm}\left(\boldsymbol{R}_{i}\right)e^{i\boldsymbol{Q}\cdot\boldsymbol{R}_{i}}\int_{\mathrm{unit\,cell}}d\boldsymbol{r}u_{v,\boldsymbol{k}-\frac{\boldsymbol{Q}}{2}}^{*}\left(\boldsymbol{r}\right)\boldsymbol{\hat{\varepsilon}}_{\pm}\cdot\boldsymbol{\nabla}u_{c,\boldsymbol{k}+\frac{\boldsymbol{Q}}{2}}\left(\boldsymbol{r}\right)\nonumber\\
 & = & \frac{\hbar}{m}A_{\pm}\left(\boldsymbol{Q}\right)\boldsymbol{\hat{\varepsilon}}_{\pm}\cdot\boldsymbol{p}_{\boldsymbol{k}-\frac{\boldsymbol{Q}}{2},\boldsymbol{k}+\frac{\boldsymbol{Q}}{2}}^{\mathrm{vc}}\nonumber\\
 & \approx & \frac{\hbar}{m}A_{\pm}\left(\boldsymbol{Q}\right)p_{\pm}^{\mathrm{vc}}\left(\boldsymbol{k}\right),
\end{eqnarray}
where we have assumed that the vector potential $\boldsymbol{\mathcal{A}}$
varies smoothly within a unit cell in the second line, with $\boldsymbol{R}_{i}$
being the $i$-th lattice site. The general vector potential $\boldsymbol{\mathcal{A}}=\boldsymbol{\hat{\varepsilon}}_+A_+(\boldsymbol{R})+
\boldsymbol{\hat{\varepsilon}}_-A_-(\boldsymbol{R})$,
with $\boldsymbol{\hat{\varepsilon}}_\pm$ being the unit vector of $\sigma^\pm$ polarization. $A_\pm\left(\boldsymbol{Q}\right)$
is the Fourier transform of $A_\pm\left(\boldsymbol{R}\right)$. $\boldsymbol{p}_{\boldsymbol{k}-\frac{\boldsymbol{Q}}{2},\boldsymbol{k}+\frac{\boldsymbol{Q}}{2}}^{\mathrm{vc}}$
is the transition matrix element from valence to conduction band.
Here we have made the approximation $\boldsymbol{p}_{\boldsymbol{k}-\frac{\boldsymbol{Q}}{2},\boldsymbol{k}+\frac{\boldsymbol{Q}}{2}}^{\mathrm{vc}}
\approx\boldsymbol{p}_{\pm}^{\mathrm{vc}}({\boldsymbol{k}})$. Consequently,
\begin{eqnarray}
\mathcal{T}_\pm & \approx & \frac{\hbar}{m S^2}\iint d\boldsymbol{R}d\boldsymbol{r}\iint d\boldsymbol{Q}d\boldsymbol{k}\Psi_{\pm}^{*}(\boldsymbol{R})\Phi^{*}(\boldsymbol{r})e^{i\left(\boldsymbol{Q}\cdot\boldsymbol{R}+\boldsymbol{k}\cdot\boldsymbol{r}\right)}A_{\pm}\left(\boldsymbol{Q}\right)p_{\pm}^{\mathrm{vc}}\left(\boldsymbol{k}\right)\nonumber\\
 & = & \frac{\hbar}{m S^2}\iint d\boldsymbol{R}d\boldsymbol{r}\Psi_{\pm}^{*}(\boldsymbol{R})A_{\pm}\left(\boldsymbol{R}\right)p_{\pm}^{\mathrm{vc}}\left(\boldsymbol{r}\right)\Phi^{*}(\boldsymbol{r})\nonumber\\
 & = & \frac{\hbar}{m S^2}\int d\boldsymbol{R}\Psi_{\pm}^{*}(\boldsymbol{R})A_{\pm}\left(\boldsymbol{R}\right)\mathcal{D}_{\pm}\nonumber\\
 & = & \mathcal{N}\int d\boldsymbol{R}A_\pm(\boldsymbol{R})\Psi_{\pm}^{*}(\boldsymbol{R}),
\end{eqnarray}
where $\mathcal{D}_{\pm}=\int d\boldsymbol{r}p_{\pm}^{\mathrm{vc}}\left(\boldsymbol{r}\right)\Phi^{*}(\boldsymbol{r})$.
If $p_{\pm}^{\mathrm{vc}}\left(\boldsymbol{k}\right)$ is a constant,
$p_{\pm}^{\mathrm{vc}}\left(\boldsymbol{r}\right)\propto\delta\left(\boldsymbol{r}\right)$
and the transition dipole $\mathcal{D}_{\pm}\propto\Phi^{*}(0)$,
which is nonzero only for $s$-wave exciton.

\section{III. Moir\'e intralayer excitons in rotationally aligned WS$\mathbf{e}_2$/WS$_2$ heterobilayer}

Lattice reconstruction usually occurs for bilayer TMDs with large moir\'e size. For rotationally aligned WSe$_2$/WS$_2$ heterobilayer, it is pointed out
that a new type of exciton, i.e., charge transfer exciton exists \cite{naik_Intralayer_2022}, whose description is
beyond the scope of continuum model \cite{wu_Topological_2017}. However, modulated Wannier excitons still exist,
and the predictions based on continuum model agree with first-principles calculations \cite{naik_Intralayer_2022}. 

\begin{figure}
\centering
\includegraphics[width=0.6\linewidth]{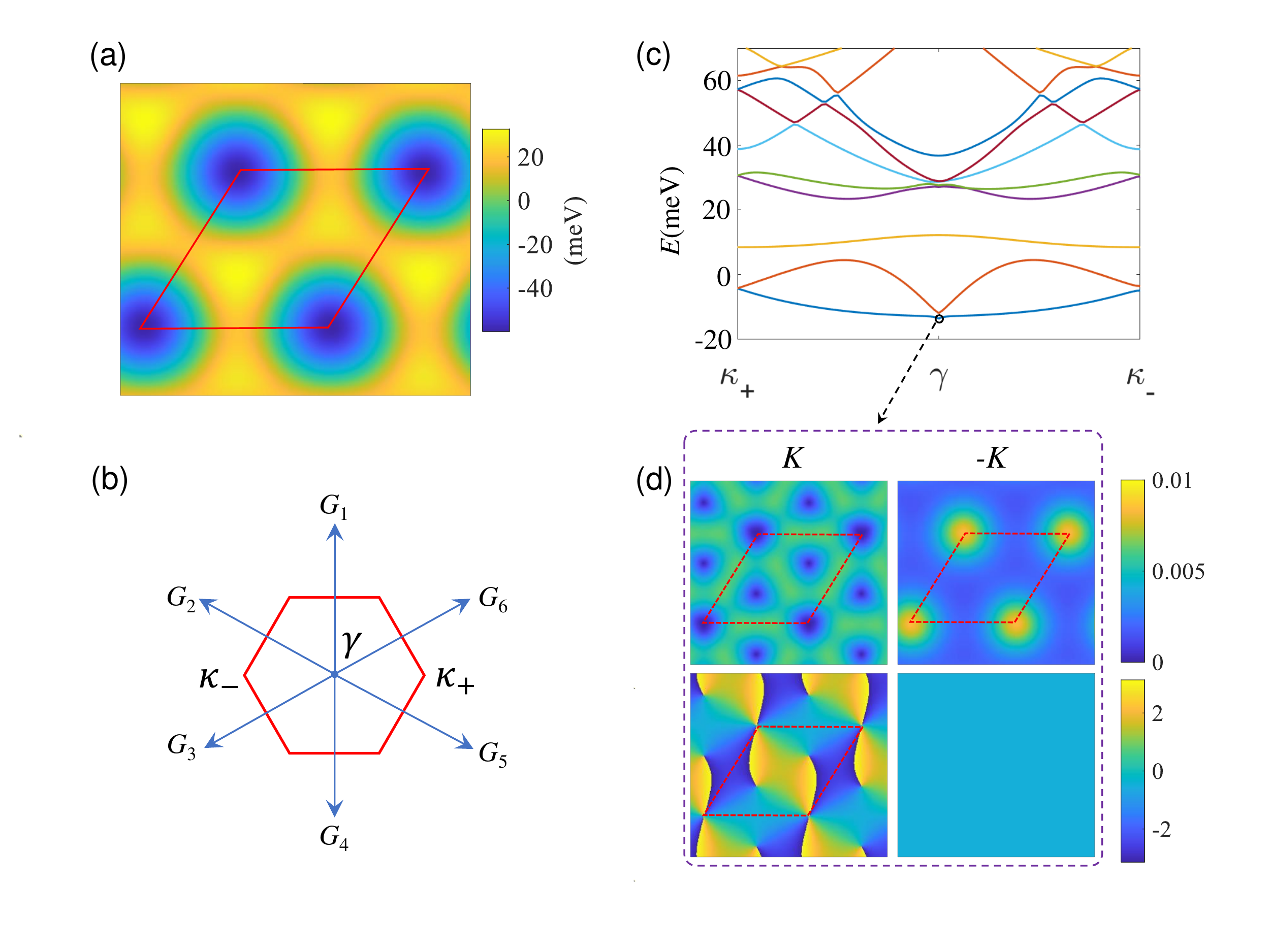}
\caption{(a) Moir\'e potential of rotationally aligned WSe$_2$/WS$_2$ heterobilayer. The moir\'e period is $8$ nm.
(b) The moir\'e exciton Brillouin zone and six reciprocal lattice vectors of moir\'e potential in (a).
(c) Exciton minibands in the moir\'e potential of (a), with a weak magnetic field ($\delta=1$ meV) breaking the degeneracy at $\gamma$ point.
The exciton Bloch wave function at ground state denoted by black circle is shown in (d),
with amplitude (phase) in upper (lower) row.
 \label{figS1}}
\end{figure}

Here we take the rotationally aligned WSe$_2$/WS$_2$ heterobilayer as an example, and
 focus on the Wannier excitons trapped by moir\'e potentials, in which case
the continuum model still works. Specifically, we model the moir\'e potential by 
$V_{\mathrm{moir\acute{e}}}=\sum_{j=1}^6 V_j \exp(i\boldsymbol{G}_j\cdot\boldsymbol{R})$,
where $\boldsymbol{G}_j$ is the moir\'e reciprocal lattice vector, and $V_j=V\exp(i\phi\times(-1)^j)$ characterizes
the strength of moir\'e potential. 
We choose $V=-10$ meV, $\phi=-3^{\circ}$, 
consistent with first-principles calculations \cite{naik_Intralayer_2022}, and $\boldsymbol{G}_1=(0,4\pi/\sqrt{3}a_m)$ with moir\'e period $a_m=8$ nm.
The other $\boldsymbol{G}_j$ is obtained by consecutive $\pi/3$ rotations shown in Fig. \ref{figS1}(b). 
 The exciton moir\'e bands by solving the Hamiltonian $\hat{H}_0+V_{\mathrm{moir\acute{e}}}$ are shown
in Fig. \ref{figS1}(c), where the degeneracy at $\gamma$ point is broken by a weak magnetic field. The Bloch wave function
at $\gamma$ point is also shown, with the phase winding around moir\'e potential minima
similar to state $\{|K,-2\rangle,|-K,0\rangle\}$.
However, the moir\'e size is not large enough, and therefore there is significant hybridization of exciton wave packets trapped at neighbouring potential minima,
which is unfavourable for the realization of single photon emitters. Since the moir\'e period here is probably already among the largest
one which supports a continuously varying superlattice potential,
lattice reconstruction for large moir\'e size poses an intrinsic obstacle to the realization of single emitter array based on
moir\'e intralayer excitons, which motivates us to propose another system based on moir\'e trions in twisted multilayer hBN substrate
in the main text.